\documentstyle[11pt,newpasp,psfig,subfigure,twoside]{article}
\index{pulsars!J0218+4232}
\index{pulsars!B1957+20}
\index{Giant pulses}

\def\edcomment#1{\iffalse\marginpar{\raggedright\sl#1\/}\else\relax\fi}
\reversemarginpar

\begin{document}

\title{Giant pulses in Millisecond Pulsars}

\author{B. ~C. ~Joshi}
\affil{National Center for Radio Astrophysics, Ganeshkhind, Pune, India}

\author{M. ~Kramer, A. ~G. ~Lyne, M. ~Mclaughlin}
\affil{Jodrell Bank Observatory, University of Manchester, Macclesfield, UK}

\author{I. ~H. ~Stairs}
\affil{Dept. of Physics and Astronomy, University of British Columbia, 6224 
Agricultural Road, Vancouver, B.C. V6T 1Z1 Canada}

\begin{abstract}
Giant pulses (GPs), occasional individual pulses with an 
intensity 100 times the average intensity, have been detected in 
four pulsars todate. Their origin is not well understood, but 
studies suggest a connection between the strength of magnetic 
field at the light cylinder $B_{lc}$ and the existence of GPs. 
Here, we report on detection of significant Large Amplitude Pulses 
(LAPs) in two more pulsars with high values of $B_{lc}$, 
PSRs J0218+4232 and B1957+20, observed using 
Giant Meterwave Radio Telescope (GMRT).
\end{abstract}

\section{Introduction}

Giant pulses (GPs) have been reported in four pulsars (B0531+21, 
B1937+21, B1821$-$24 and B0540$-$69) todate (Staelin \& Reifenstein ~1968; 
Lundgren et~al. ~1995; Cognard et~al. ~1996; Romani \& Johnston ~2001; 
Johnston \& Romani ~2003). Three of these pulsars are millisecond pulsars 
(MSPs) and also show strongly pulsed hard X-ray profiles 
(Takahashi et~al. ~2001). The radio GPs occur in a narrow phase window 
close to the high energy non-thermal pulse indicating a common 
magnetospheric origin. All these pulsars 
have a high value of $B_{lc}$. We have used GMRT to search for 
GPs in candidate MSPs with a non-thermal high energy emission 
and a range of $B_{lc}$ and report detection of such pulses in 
two more pulsars, PSR J0218+4232 and B1957+20. 

We obtained about 3600 s of data on each pulsar using 20 to 
22 GMRT antennae in an incoherent mode at 610 MHz with 16 MHz bandwidth. 
The expected RMS noise in the above configuration of GMRT for a 
sampling time of 258 $\mu s$ is about 1 Jy.
The data in two subbands (8 MHz each) were dedispersed to a common sky 
frequency (610 MHz). The periods with a peak greater than 3.5 times RMS 
in both bands at the same sample were identified as Large amplitude 
pulses (LAPs). This procedure disperses any narrow interference spike 
discriminating against interference. A number of marginal LAPs, i.e. 
pulses with a peak between 3.0 to 3.5 times RMS, were also identified.

\begin{figure}
\centering
\subfigure{\mbox{\psfig{file=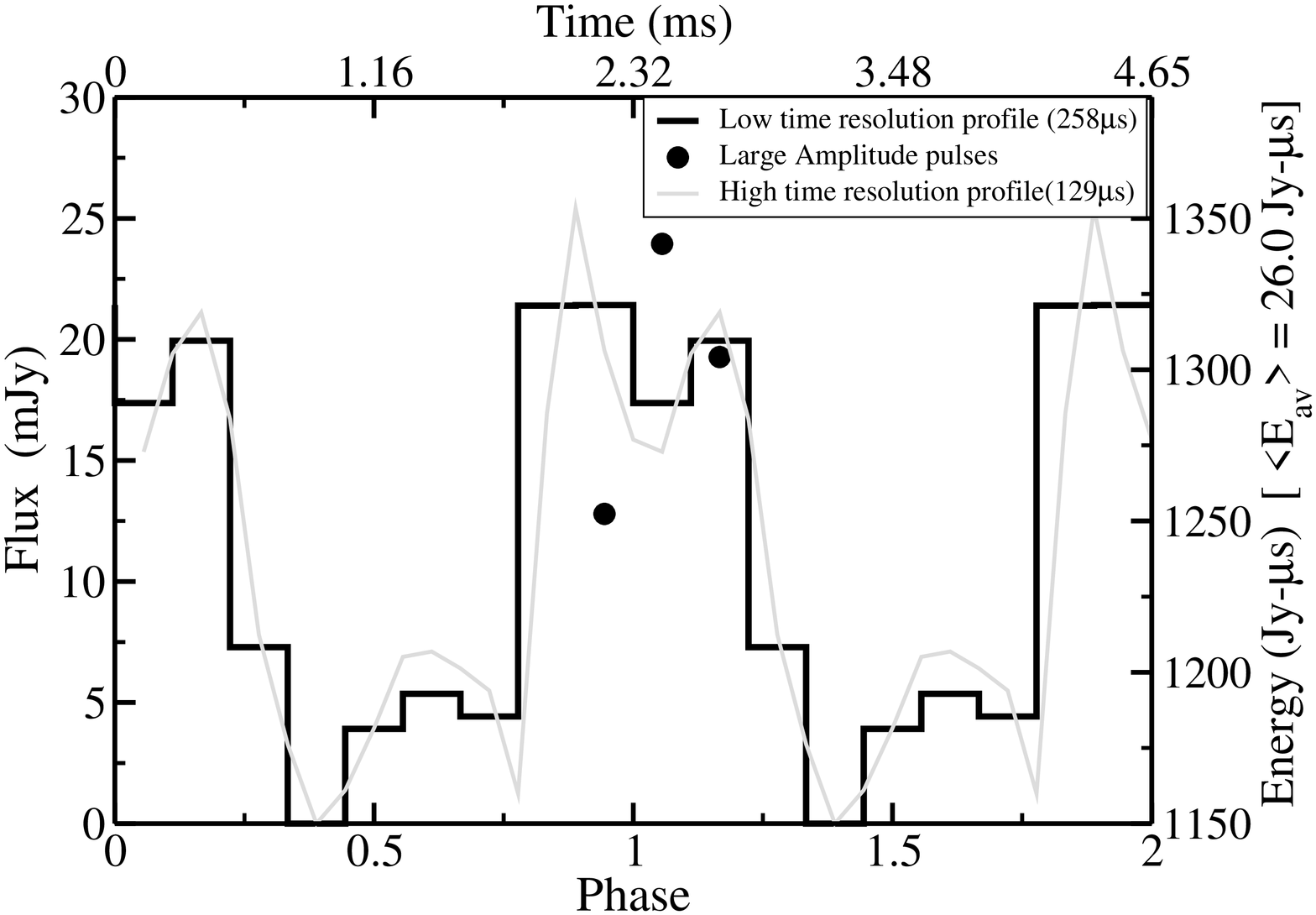,angle=270,width=2.5in}}}\quad
\subfigure{\mbox{\psfig{file=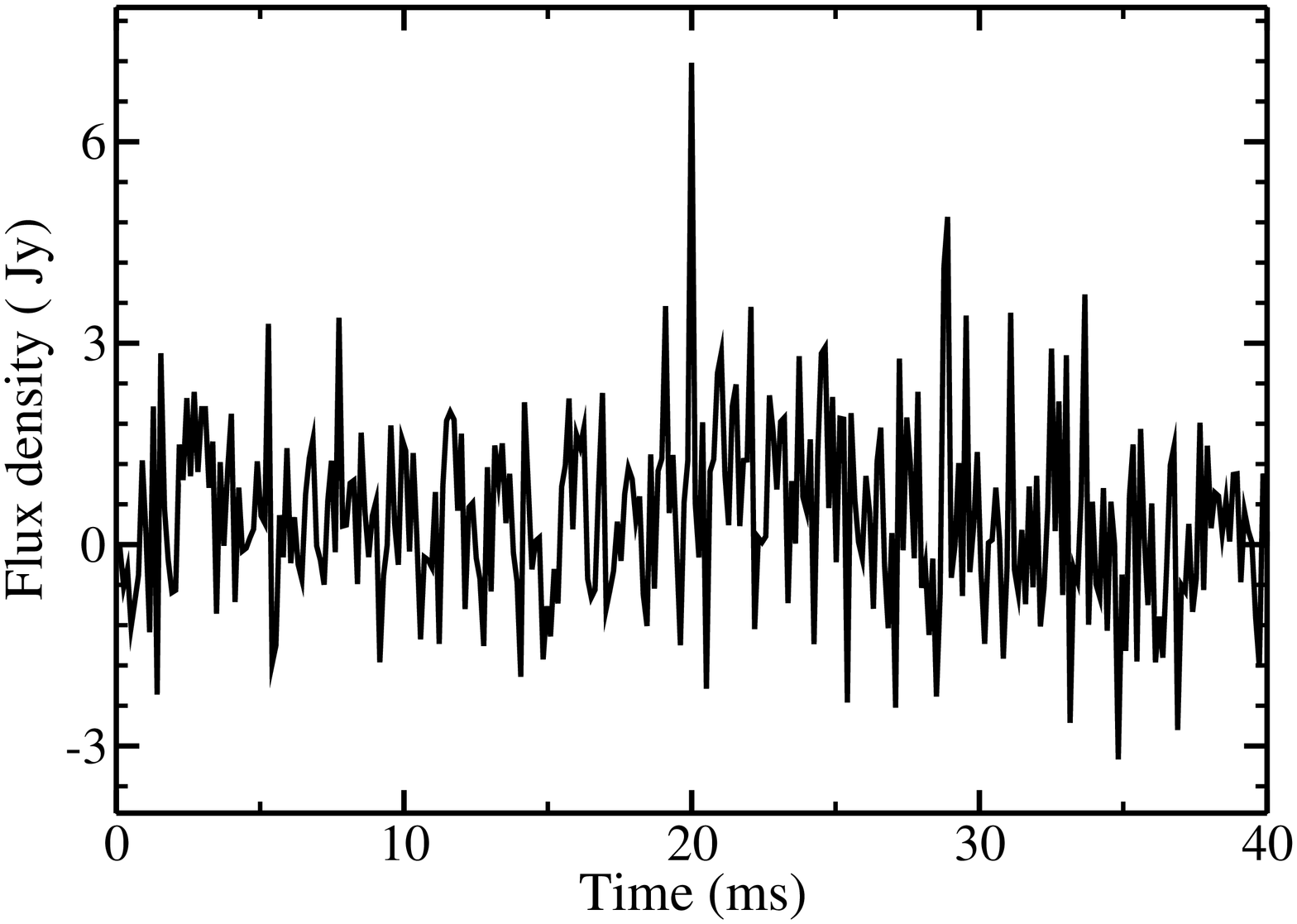,angle=270,width=2.5in}}}
\caption{LAPs in PSR J0218+4232 and B1957+20. a) The left panel shows the 
baseline subtracted radio integrated profile for PSR J0218+4232 
with the LAPs marked with the filled 
circles at the phase of their occurrence. The energy in the LAP is labeled 
on the right. b) The right panel shows the GP detected in PSR B1957+20 
after removing the off-pulse mean. }
\end{figure}


\section{Results}

We searched 2.2 million periods for PSR J0218+4232 
($B_{lc} \sim$ 3.2 $\times 10^5$ G) and found three significant LAPs. 
Figure 1a shows the integrated profile for 
this pulsar and the detected LAPs are marked with filled circles.
The largest of these had an intensity 51 times the mean intensity 
(Intensity of LAP, $I_{lap} \sim 1341$ Jy-$\mu$s). 
Our data also consisted of 9 marginal LAPs, 7 of which occur between 
phase 0.89 - 1.2, which is the phase interval corresponding to one of the 
high energy peaks.
We searched about 1 million periods for PSR B1957+20 
($B_{lc} \sim$ 3.8 $\times 10^5$ G) and found one significant 
LAP, shown in Figure 1b, with an intensity 129 times the mean intensity 
($I_{lap} \sim 925$ Jy-$\mu$s). In addition, we also detected 
5 marginal LAPs.

PSRs B1957+20 and J0218+4232 have the fourth and sixth 
highest values of $B_{lc}$ respectively of known radio pulsars. Hence, 
these new 
detections support a connection between the magnitude of $B_{lc}$ 
and the existence of GPs and LAPs. The data for pulsars with $B_{lc}$ 
marginally below $10^5$ G are being analyzed currently. 


\begin{references}
\reference  {Cognard}, ~I. {\rm et~al.} 1996, ApJ, 457, 81
\reference  Johnston, ~S., \& Romani, ~R.~W. 2003, private communication
\reference  Lundgren, ~S.~C. {\rm et~al.}  1995, ApJ, 453, 433
\reference  Romani, ~R. ~W., \& Johnston, ~S. 2001, ApJ, 557, L93
\reference  Staelin, ~D.~H., \& Reifenstein, ~E.~C. 1968, Science, 162, 1481
\reference  {Takahashi}, ~M. {\rm et~al.} 2001, ApJ, 554, 316
\end{references}


\end{document}